\documentclass[preprint,12pt]{elsarticle}

\usepackage{amssymb}

\journal{ Physics Letters  B}

\begin{document}
\def\sigmatot{$\sigma_{total}$}
\def\sigtot  {\mbox{$\sigma_{\rm tot}^{\rm pp/ p \bar p}$}}
\def\sigtotpp  {\mbox{$\sigma_{\rm tot}^{\rm pp}$}}
\def\rs{\mbox{$\sqrt{s}$}}
\def\pbarp{\mbox{$\rm  \bar{p}p$}}
\def\ptmin{\mbox{$p_{tmin}$}}
\newcommand{\ra}{\rightarrow}
\newcommand{\ba}{\begin{array}}
\newcommand{\ea}{\end{array}}
\newcommand{\beqa}{\begin{eqnarray}}
\newcommand{\eeqa}{\end{eqnarray}}
\newcommand{\be}{\begin{equation}}
\newcommand{\ee}{\end{equation}}

\begin{frontmatter}


 \title{Soft Gluon $k_t$-Resummation and 
the Froissart bound}



\author[1]{Agnes Grau}
\ead{igrau@ugr.es}
\author[2]{Rohini M. Godbole}
\ead{rohini@cts.iisc.ernet.in}
\author[3]{Giulia Pancheri}
\ead{pancheri@lnf.infn.it}
\author[4]{Yogendra N. Srivastava}
\ead{yogendra.srivastava@pg.infn.it}
\address[1]{Departamento de Fisica Teorica y del Cosmos, Universidad de
 Granada, Spain}
\address[2]{Centre for High Energy Physics, Indian Institute of Science,
Bangalore,
560012,India}
\address[3]{INFN Frascati  National Laboratories, P.O. Box 13, Frascati I00044,
    Italy}
\address[4]{Physics Department and INFN, University of Perugia,
            Perugia, Italy}
\begin{abstract}
We study soft gluon $k_t$-resummation and the relevance of  InfraRed (IR)  
gluons for the energy dependence of  total hadronic cross-sections.  
 In our  model,
consistency 
with the Froissart bound
   is directly related to the
  ansatz that the IR  behaviour of  the QCD
coupling constant follows an  inverse  power law.
\end{abstract}

\begin{keyword}
 Froissart bound \sep  QCD \sep  total cross-section \sep resummation

\PACS 13.85.Lg \sep 12.38.Cy


\end{keyword}

\end{frontmatter}

 \section{Introduction}
  In this paper, we  discuss  soft gluon $k_t$-resummation in the 
InfraRed (IR) region, with the aim to connect it to the energy dependence of 
the total hadronic cross-section in the high energy region. 
 We shall make use of a model \cite{GGPS} for total cross-sections, 
which incorporates in an eikonal formulation such QCD inputs as mini-jets 
and soft gluon  $k_t$-resummation. This model has been successfully applied 
both to proton and photon processes: our aim 
 here is to describe 
its physical content, without explicit reference to data fitting, 
 and explain 
 how the model incorporates a taming effect on the rapidly rising QCD 
cross-sections,  thus
inducing a more temperate rise of the total cross-section.

The energy behaviour of the total hadronic cross-section has been  the focus
  of
both theoretical and experimental enquiries for a long time. Fits inspired
by theoretical arguments have been the subject of many debates.
 A  traditional 
 Regge-Pomeron type fit \cite{DL} such  as
\begin{equation}
\label{DL}
\sigma_{total}=X s^{-\eta}+ Ys^{\epsilon},
\end{equation}
 with $\eta,\epsilon >0$, presents the difficulty  of not
agreeing with the Froissart-Lukaszuk-Martin  bound
 \cite{froissart,martinold,luk} 
\begin{equation}
\label{flimit}
\sigma_{total} \le {{\pi}\over{m_\pi^2}} \ln^2( s/s_0).
\end{equation}
  However,
with $\sqrt{s_0}\simeq {\cal O}(1\ GeV)$, 
 the large constant factor  at the r.h.s  of Eq.~(\ref{flimit})
 makes 
possible for  Eq.~(\ref{DL})  to be valid  only in the present region, and 
the observed behaviour to be not yet 
asymptotic. 
Still,  phenomenological
reasons run against the  validity of the  Regge-Pomeron type expression  
with a universal value for $ \epsilon$,   since
LEP data on photon-photon
  indicate  that the power, with which
$\sigma_{total}^{\gamma \gamma}$ rises, differs from that in $pp/p{\bar p}$
\cite{albert}.


 Currently, many models focus on QCD perturbative processes to drive the rise 
with energy, with QCD inspired models on the one hand \cite{BGHP,Luna} and  
approaches based on Reggeon calculus \cite{levin,khoze} on the other. 
However, it is still being debated how to implement this dynamics and 
simultaneouly describe both the early rise, which starts  
for $\sqrt{s}$ between $10\div 20\ GeV\ $ and $ 50\div 60\ GeV$, and the 
subsequent levelling off at  higher energies \cite{PDG}.

In this paper we shall  show how both of the above features are present in 
our model for the total cross-section \cite{GGPS}
  and  relate the rate, at which the cross-section asymptotically rises,
  to the IR  limit of soft $k_t$-resummation, thus
 linking directly the rise of the total cross-section to the 
 IR region of QCD.
 We  start  by  recalling in Sect.~\ref{sec:resum}  
 some  features of resummation
  and  present, in Sect.  ~\ref{alphas},
   our 
 proposal for handling 
the ultra soft gluon emissions which affect scattering at the very large
impact parameter values,    relevant to  
total
cross-section calculations.  After a brief discussion of our model in  
Sect.~\ref{sec:BNmodel},
  we show in Sect.~\ref{sec:derivation}
 the connection between the Froissart bound and our proposed eikonalized 
mini-jet model with ultra soft gluons.  Finally, in 
Sect.~\ref{sec:scales}
we discuss both the various energy scales and the constants involved in the 
model.  We find that 
 resummation  of soft gluons emitted with very small transverse momentum 
$k_t$ introduces a new  energy scale, which 
  modifies the constant in front of the asymptotic   limit derived in  our 
model for the total cross-section.

\section{\label{sec:resum} Resummation and the IR limit}
 The high energy behaviour of the total cross-section depends on the 
properties of the scattering amplitude at large values of the impact parameter
 $b$ in the plane perpendicular to the scattering. Recent attempts to study 
this large-$b$ behaviour have focused on relating Yang-Mills theories to 
string theories through the AdS/CFT correspondance \cite{tan,bpst,domokos}. 
Our approach to the large $b$-limit of  the impact picture in the eikonal 
representation \cite{bsw}  
 is of a more phenomenological nature:
within such a picture, we exploit 
  soft  gluon $k_t$-resummation 
 in the IR region 
to describe matter distribution inside the hadrons as they  engage in hard 
scattering and their parton constituents ``see''   each other.  
We have already supplied phenomenological evidence for 
the applicability of our  model to high energy scattering \cite{GGPS}. Here 
we discuss some features of soft gluon $k_t$-resummation 
which bear on  the asymptotic form of the total cross-section. 

We start by recalling some properties of 
soft photon
 resummation. In QED,  the general expression for 
 soft photon resummation in the energy-momentum variable
$ K_\mu$  can be obtained order by order in perturbation 
theory \cite{jr,yfs}
as 
\begin{equation}
d^4P( K)=d^4  K
\int {{d^4x} \over {(2 \pi)^4}}
e^{i{ K \cdot x} -h( x,E)}
\label{d4pk}
\end{equation}
where $d^4P(K)$ is the probability for an overall 4-momentum $K_\mu$ 
escaping detection,  
\begin{equation}
h(x,E)=\int_0^E d^3{\bar n}(k)[1-e^{-ik\cdot x}]
\label{hdE}
\end{equation} 
with $d^3{\bar n}(k)$ being the single soft photon differential spectrum,
and $E$ 
the maximum energy allowed for  single photon emission.

  Eq.~(\ref{d4pk}) leads to the well known, power-like,  form of the 
energy distribution \cite{lomon}.
This is not possible  for the momentum distribution, but  
  it is also not necessary, since the first order expression in 
$\alpha_{QED}$ is adequate.
On the other hand, for  strong interactions we require  the 
 resummed, transverse momentum distribution, namely
 \begin{equation}
 d^2P({\bf K_\perp})=d^2{\bf  K_\perp} {{1} \over{(2\pi)^2}}\int
d^2 {\bf b}\ e^{-i{\bf K_\perp\cdot b} -h( b,E)}
\label{d2p}
 \end{equation}
 with
\begin{equation}
 h(b,E)=\int d^3{\bar n}(k) [1-e^{i \bf k_\perp\cdot b}]
 \label{hdb}
\end{equation}
 For large   transverse momentum values, 
 by neglecting the second term and using a  constant cut-off  as lower limit 
of integration, the above expression coincides with the   Sudakov 
form factor \cite{sudakov}.
 
Eq.~(\ref{d2p}) has been applied in  QCD 
\cite{ddt,pp,halzenscott,altarelli,curci}
with
\begin{equation}
h( b,E) =  \frac{16}{3}\int^E 
 {{ \alpha_s(k_t^2) }\over{\pi}}{{d
 k_t}\over{k_t}}\ln{{2E}\over{k_t}}[1-J_0(k_tb)]
 \label{hbq}
\end{equation}
but its   use 
is complicated by
our ignorance of the IR behaviour of the theory.
 To overcome the difficulty arising from the IR region, the function $h(b,E)$, which  describes the relative  transverse momentum distribution induced by soft gluon emission from  a pair of,
initially collinear, colliding partons  at LO, 
is split into
\begin{equation}
\label{h1}
h(b,E) = c_0(\mu,b,E)+ \Delta h(b,E),
\end{equation}
where
\begin{equation}
\label{h2}
\Delta h(b,E) =
 \frac{16}{3} \int_\mu^E {\alpha_s(k_t^2)\over{\pi}}[1- J_o(bk_t)]
 {{dk_t}\over{k_t}}
  \ln {
  {{2E}\over{k_t}}}.
\end{equation}
 Since the integral in $\Delta h(b,E)$
now  extends down to a   scale $\mu \neq 0$, 
for $\mu> \Lambda_{QCD}$ one can use the asymptotic freedom expression 
for $\alpha_s(k_t^2)$.  Furthermore, having excluded the zero momentum 
region from the integration,  
 $J_o(bk_t)$ is  assumed  to oscillate to zero and    neglected.
The   integrand 
in Eq.~(\ref{h2}) is now independent
of $b$ and   the integral can be performed.
In the range $1/E < b < 1/\Lambda$, the effective $h_{eff}(b,E)$ is 
obtained by setting $\mu = 1/b$ \cite{pp}. This choice of the scale   
introduces a cut-off in impact parameter space which is stronger than 
any power, since
the radiation function, for $N_f=4$,  is now \cite{pp}
 \begin{equation}
e^{-h_{eff}(b,E)} =
 \big{[}
{{
\ln(1/b^2\Lambda^2)
}\over{
\ln(E^2/\Lambda^2)
}}
\big{]}^{(16/25)\ln(E^2/\Lambda^2)}
\label{PP}
\end{equation}
The remaining  $b$-dependent term, namely 
$exp[-c_0(\mu,b,E)]$,
is dropped, a reasonable approximation if one  assumes that there is no 
physical singularity in the range of integration
 $0\le k_t \le 1/b$.  This contribution however reappears as an energy 
independent  smearing function which reproduces phenomenologically 
 the effects of an intrinsic  transverse momentum of partons.
For most applications, this may be a good approximation. 
However, when  the integration in impact parameter space extends to very 
large-$b$ values, as  is the case for the calculation of total 
cross-sections, the IR region may be  important and  the possibility of  a 
physical singularity for $\alpha_s$ in the IR region becomes relevant. 
It is this possibility,   which we exploit in studying scattering in the very 
large impact parameter region, $b\rightarrow \infty$.

 \section{ A proposal for the IR limit in the soft gluon 
integral\label{alphas}}
  In this Section we discuss  a phenomenological expression for the coupling 
of  ultra soft gluons    to the emitting quarks, and compare the resulting 
large-$b$ behaviour with the one discussed in the previous  section.
 
 Our choice for the IR behaviour of $\alpha_s(Q^2)$  used in obtaining  a
  quantitative description of the  distribution in Eq. (\ref{hdb}),
  is inspired by the   Richardson potential
 for quarkonium bound states \cite{RICHARDSON}, as we have proposed in   a 
number of related
applications \cite{DY}.
Assume a confining potential (in momentum space) given by   the one gluon
exchange term
\begin{equation}
{\tilde V}(Q) = K ({{\alpha_s(Q^2)}\over{Q^2}}),
\end{equation}
where K is a constant, calculable from the asymptotic form of 
$\alpha_s(Q^2)$. Let
us choose for $Q^2<<\Lambda^2$ the simple form
\begin{equation}
\alpha_s(Q^2) = {{B�}\over{ (Q^2/\Lambda^2)^{p}}},
\end{equation}
(with B a constant), so that ${\tilde V}(Q)$ for small Q goes as
\begin{equation}
\label{potential}
{\tilde V}(Q) \rightarrow  Q^{-2(1 +p)}.
\end{equation}
For the potential, in coordinate space, $
V(r) =\ \int d^3Q/(2\pi)^3  e^{i {\bf Q.r}} {\tilde V}(Q)$,
Eq.(\ref{potential}) implies
\begin{equation}
V(r) \rightarrow (1/r)^3 \cdot r^{(2 + 2p)} \sim  C\ r^{(2p - 1)},
\end{equation}
for large r (C is another constant). A simple check is that for $p$
equal to zero, the usual Coulomb potential is regained. Notice that for
a potential rising with r, one needs $p > 1/2$. Thus, for $1/2< p< 1$,
this corresponds to a confining potential rising less than linearly
with the interquark distance $r $, 
while a  value of $p=1$ coincides with the IR limit of the Richardson's potential and is also found in a number of applications to potential
estimates of quarkonium properties \cite{yndurain}.

Then,  again following Richardson's argument,  we connect our IR limit for $\alpha_s(Q^2)$ to the asymptotic freedom region 
  using the  phenomenological  expression:
\begin{equation}
\label{alphapheno}
\alpha_s(k_t^2)={{12 \pi }\over{(33-2N_f)}}{{p}\over{\ln[1+p({{k_t^2}
\over{\Lambda^2}})^{p}]}}
\end{equation}
which coincides with the usual one-loop formula  for
 values of $k_t>>\Lambda$, while going to a singular
 limit for small $k_t$,  and 
generalizes Richardson's ansatz to values of $p\le 1$. The range $p < 1$ has 
an important advantage,
i.e., it allows the integration in Eq.(\ref{hdb}) to converge for all
values of $
k_t=|k_\perp |$. 
Using Eq.~(\ref{alphapheno}), one can study the behaviour of $h(b,E)$ for the 
very large-$b$ values  which enter the total cross-section calculation and 
recover the perturbative calculation as well.  The behaviour of  
$h(b,E)$ in   various regions in $b$-space was discussed 
in \cite{our99},  both  for a singular and a frozen $\alpha_s$, 
namely one whose IR limit is a constant. 
 There we saw that, for the singular $
 \alpha_s$ case,  the following is a good analytical approximation in 
the very large-$b$ region:
 \begin{eqnarray}
\label{halphas3}
h(b,M,)=
{{2c_F}\over{\pi}}\left[ {\bar b} {{b^2\Lambda^{2p}}\over{2}}
\int_0^{{1}\over{b}}{{dk}\over{k^{2p-1}}} \ln {{2M}\over{k}}+
2 {\bar b} \Lambda^{2p}\int_{{1}\over{b}}^{N_p\Lambda} {{dk}\over{k^{2p+1}}}
\ln {{M}\over{k}}+{\bar b} \int_{N_p\Lambda}^M {{dk}\over{k}} 
{{\ln{{M}\over{k}}}\over{\ln {{k}\over{\Lambda}}}}\right] &&\nonumber \\
 ={{2c_F}\over{\pi}} \Biggl [ {{{\bar b}}\over{8(1-p)}} (b^2\Lambda^2)^p
\left[ 2\ln(2Mb)+{{1}\over{1-p}}\right] +
 {{\bar b}\over{2p}}(b^2\Lambda^2)^p \left[2\ln(Mb)-{{1}\over{p}}\right]
+\nonumber &&
\\{{\bar b}\over{2pN_p^{2p}}}\left[-2\ln{{M}\over{\Lambda N_p}}+{{1}\over{p}}
\right] + 
 {\bar b} \ln {{M}\over{\Lambda}}\left[\ln {{\ln{{M}\over{\Lambda}}}\over
{\ln{N_p}}}-1+{{\ln{N_p}}\over{\ln{{M}\over{\Lambda}}}} \right] \Biggr ]
\end{eqnarray}
 This approximation   is valid in the region  $b>1/(N_p\Lambda)>1/M$,
with $N_p=(1/p)^{1/2p}$, $c_F=4/3$ for emission from quark 
legs and ${\bar b}=12 \pi/(33-2N_f)$. The upper limit of integration, 
$M$,   indicates  the maximum  allowed  transverse momentum, 
to be determined
 by  the kinematics of single gluon emission as in \cite{GRECO}.  
The above  expression  exhibits the sharp cut-off at large-$b$ values which 
we shall exploit to study the very large energy behaviour of our model.
On the other hand, the possibility that $\alpha_s$ becomes constant in the IR
\cite{pp,halzenscott,altarelli}
in   the same  large $b$-limit 
 leads to
\begin{equation}
h(b,M,\Lambda)=(constant) \ln(2Mb) + \ double \ \ logs
\end{equation}
namely no sharp cut-off in impact parameter $b$, as expected. 

\section{\label{sec:BNmodel} The Bloch-Nordsieck(BN) model for 
the total cross-section}
Our model for the total cross-section \cite{GGPS} is a modified 
mini-jet model, in which the rise with energy  is 
driven by  perturbative parton-parton scattering \cite{halzen}, 
tempered by  an energy dependent acollinearity effect.  
This effect  is due  to  $k_t$-resummation of soft gluon emission from 
the initial state, hereafter referred to as 
{\it soft gluon $k_t$-emission}.  The  emphasis on resummation, first 
introduced for electron scattering by Bloch and Nordsieck \cite{BN}, 
gives the model its name. The model  is built through the eikonal 
representation  in impact parameter space,  so as to satisfy unitarity, and  
 allows to implement multiple parton scattering and to restore a finite size 
of the interaction through the impact parameter distribution in the 
scattering hadrons.  The details of the model can be found in 
\cite{GGPS,our99,corsetti}, here we shall  recall   some aspects relevant to 
its asymptotic energy behaviour.

 In 
  hadron-hadron scattering at a c.m. energy $\sqrt{s}$,
   unitarity allows to write a simple model for the total cross-section, 
namely
\begin{equation}
\sigma_{total}=
2
\int d^2{\bf b} 
[
1-e^{
-{\cal I} m \chi(b,s)
}cos\Re e\chi(b,s)
] \approx 
2
\int d^2{\bf b} 
[
1-e^{
-{\bar n}(b,s)/2
}
]
\end{equation}
 where the approximation on the r.h.s is obtained by neglecting the real 
part of the eikonal function (at the hadronic level,   an acceptable approximation in the high energy limit) and 
$2{\cal I} m \chi(b,s)={\bar n}(b,s)$. The latter 
follows from 
a  semiclassical argument  relating 
  $\sigma_{inelastic}$ to a sum of a Poisson distributed independent, single and multiple collisions. 


 We use 
perturbative QCD  to calculate   the cross-section in order to obtain 
 the average number of inelastic collisions.  
While implementing  the  QCD calculation, albeit approximate, 
we distinguish between 
the average number
of collisions receiving contributions from   hard  physics processes and 
those from non perturbative ones, and write ${\bar n}(b ,s)$  in the form
\begin{equation}
 {\bar n}(b,s) = n_{NP} (b,s) + n_{hard} (b,s)
\end{equation}
where the  non perturbative (NP) term parametrizes the contribution of
 all those processes for which    initial partons scatter with 
$p_t<p_{tmin}$, with $p_{tmin} $ a suitable low energy cut-off for the QCD 
parton-parton cross-section. 
 We parametrize  $n_{NP}(b,s)$, which 
 establishes the overall normalization, 
 and focus our attention 
 on  the hard term,  which is
responsible for the high-energy rise and which we expect to dominate in the  
extremely high energy limit.  We approximate this term
as
\begin{equation}
 n_{hard} (b,s) = A(b,s)\sigma _{jet} (s). 
  \label{nhard}
 \end{equation}
The QCD jet cross section drives the rise due to the
increase with energy of the number of partonic collisions. It  is calculated 
from the usual perturbative QCD expression,
with DGLAP evoluted
parton densities
and  perturbative partonic differential cross-sections.
In Eq. (\ref{nhard}), $A(b,s)$ is the overlap function  which  depends on   
the (energy dependent) spatial
distribution of partons inside the colliding hadrons,  averaged over the 
densities \cite{GGPS,corsetti}. Before discussing this function we shall 
examine the energy behaviour of the mini-jet cross-sections.

In the
$\sqrt{s}>>p_{tmin}$ limit, the major contribution to  the mini-jet cross-sections 
comes from collisions of gluons carrying 
 small momentum fractions  
  $x_{1,2}<<1$, a region  where   the relevant 
PDFs   behave approximately like powers of the momentum fraction
$x^{-J}$ with $J \sim 1.3$ \cite{Lomatch}.  This leads  to
 the 
 asymptotic high-energy expression for
$\sigma_{jet}$
\begin{equation}
\sigma _{jet}  \propto \frac{1}{{p_{t\min }^2 }}\left[
{\frac{s}{{4p_{t\min }^2 }}} \right]^{J - 1} 
\label{minijetasympt}
\end{equation}
 where the dominant term is   a power of $s$. 
 Fits to the mini-jet cross-sections, obtained with different PDF sets 
\cite{mpi08} confirm the value 
$\varepsilon\equiv J-1\sim0.3$.  

Such energy behaviour as in Eq.~(\ref{minijetasympt}) is at
 odds with the gentle rise of the total $pp$ and $p {\bar p}$ cross-sections 
at very high energy, 
 described rather as   $\ln {s, }\ln^2{s}$ \cite{PDG} or  $s^{0.08}$ 
power \cite{DL}.
However, we shall see that a proper implementation of other QCD processes 
can  modify  this strong rise. To do so we now examine the energy behaviour 
of the next component of our BN model, the impact parameter distribution.

We  have identified  soft gluon $k_t$-emission from the colliding partons 
as the
physical effect responsible for the attenuation of the rise of the
total cross section. These soft emissions 
break collinearity between the colliding partons, diminishing the
efficiency of the scattering process. Their  number 
increases with  energy and thus their contribution remains 
important, even  at very high energy, 
  influencing matter distribution
inside the hadrons, hence changing the overlap function, which is proposed to 
be
the Fourier transform of the
previous expression for the soft gluon transverse momentum resummed 
distribution,  i.e,. we put
\begin{equation}
A_{BN}(b,s)=N \int d^2{\bf  K_{\perp}}\  e^{-i{\bf K_\perp\cdot b}}
 {{d^2P({\bf K_\perp})}\over{d^2 {\bf K_\perp}}}={{e^{-h( b,q_{max})}}\over
 {\int d^2{\bf b} \ e^{-h( b,q_{max})}
 }}=A_0(s) e^{-h( b,q_{max})}
 \label{adb}
\end{equation}
The integral in $h( b,q_{max})$
 is performed up to a value $q_{max}$, which is linked
to the maximum transverse momentum allowed by  kinematics of single gluon 
emission
 \cite{GRECO}. In principle, this parameter and the
overlap function should be calculated for each partonic
sub-process, but in  the partial factorization of Eq.(\ref{nhard}) we  use an 
average value of
$q_{max}$ obtained by considering all the sub-processes that can
happen for a given energy of the main hadronic
process\cite{our99}. The energy parameter $q_{max}$ is of the order of 
magnitude of $p_{tmin}$. For present low$-x$ behaviour of the PDFs,  in  the 
high energy limit,  $q_{max}$
is a slowly varying function of $s$, starting as $\ln{s}$, with a limiting 
behaviour which depends on the densities  
\cite{kazimierz}.
From Eqs.~(\ref{halphas3}) and (\ref{minijetasympt}) 
one can estimate the very large $s$-limit 
\begin{equation}
n_{hard} (b,s)=A_{BN}(b,s)\sigma_{jet}(s,p_{tmin})\sim A_0(s)  
e^{-h(b,q_{max})}{\sigma_1}({{s}\over{s_0}})^\varepsilon
\end{equation}
and, from this, using  the very large $b$-limit,
\begin{equation}
n_{hard} (b,s)\sim A_0(s) \sigma_1 e^{-(b{\bar \Lambda})^{2p}} 
({{s}\over{s_0}})^\varepsilon
\end{equation}
with $A_0(s)\propto 
\Lambda^2 $ and with a logarithmic dependence on $q_{max}$, i.e.  a 
very slowly varying function of $s$. We also have  
\begin{equation}
{\bar \Lambda}\equiv
 {\bar \Lambda(b,s)}=\Lambda 
 \{ 
 {{c_F{\bar b}}\over{4\pi(1-p)}}[
 \ln (2q_{max}(s)b) +{{1}\over{1-p}}]
 \}^{1/2p}
 \end{equation}
In the next section, we shall see how the   two critical exponents of our 
model, namely the power $\varepsilon$ with which the mini-jet cross-section 
increases with energy and the parameter $p$  dictating the IR behaviour of 
the QCD coupling constant, combine to obtain a rise of the total 
cross-section in agreement with the $\ln^2s$ limitation imposed by the 
Froissart bound.

\section{\label{sec:derivation} Ultra soft  gluons in the IR limit and
 the asymptotic limit of the total cross-section: the Froissart bound}
In this section we consider the very large $s$-limit of the total 
cross-section, in the approximation that all constant  (or decreasing )
 terms in the eikonal 
function can be neglected, thus studying only the QCD effects from mini-jets 
and soft $k_t$-resummation. We find a link  between the infrared behaviour of 
the ultra soft gluons and the asymptotic Froissart-like behaviour of the 
total cross-section and discuss it.

Let us consider the total cross-section in the eikonal representation at 
very large asymptotic energies. At such  large energies that 
$n_{NP}<<n_{hard}$, the total cross-section in our model \cite{GGPS} 
reads
\begin{equation}
  \sigma _T (s) \approx 2\pi \int_0^\infty  {db^2 } 
[1 - e^{ - n_{hard} (b,s)/2} ]
\end{equation}
We consider the asymptotic expression for $\sigma_{jet}$ at high
energies, which grows like  a power of $s$, 
and  $A_{BN}(b,s)$,  which was   obtained
through soft gluon resummation,  and which decreases  in $b$-space at least 
like an exponential ($1<2p<2$).
In such large-$b$, large-$s$ limit, we can write 
\begin{equation}
  n_{hard}  = 2C(s)e^{ - (b{\bar \Lambda})^{2p} }
\end{equation}
where  $2C(s) = A_0(s) \sigma _1 (s/s_0 )^\varepsilon  $. The resulting
expression for $\sigma_T$ is
\begin{equation}
  \sigma _T (s) \approx 2\pi \int_0^\infty  {db^2 } [1 - e^{ -
C(s)e^{ - (b{\bar \Lambda})^{2p} } } ]
\label{sigT}
\end{equation} 
With the variable transformation
$u=({\bar \Lambda} b)^{2p}$,
and neglecting the logarithmic $b$-dependence in ${\bar \Lambda}$ by 
putting $b=1/\Lambda$,
Eq.~(\ref{sigT}) becomes
\begin{equation}
 \sigma _T (s) \approx {{2\pi}\over{p}}{{1}\over{{\bar \Lambda}^2}} 
\int_0^\infty du u^{1/p -1}
 [1-e^{
 -C(s)
 e^{-u}
 }
 ]  
\end{equation}
Notice that, as
$s\rightarrow \infty$, $ C(s)$ also grows indefinitely as a power
law. This means that the quantity between square brackets
$I(u,s)=1-e^{-C(s)e^{-u}}$ has the limits
$I(u,s)\rightarrow 1$ at $u=0$
and $I(u,s)\rightarrow 0$ as $u=\infty$.
Calling $u_0$ the value at which $I(u_0,s)=1/2$ we then put
$I(u,s)\approx 1$ and integrate only up to $u_0$. Thus
\begin{equation}
{\bar \Lambda}^2\sigma_{T}(s)\approx (\frac {2\pi} {p})\int_0^{u_0} du u^{\frac{1-p} {p}}
=2\pi u_0^{1/p}
\end{equation}
and since, by construction
\begin{equation}
u_0=\ln[\frac {C(s)}{\ln 2}]\approx \varepsilon \ln s
\end{equation}
we finally obtain
\begin{equation}
\sigma_{T}\approx \frac {2\pi } {{\bar \Lambda}^2} 
[\varepsilon \ln \frac {s} {s_0}]^{1/p}
\label{froissart1}\end{equation}
to leading terms in $\ln s$. 
 We therefore derive the asymptotic energy dependence 
\begin{equation}
  \sigma _T  \to [\varepsilon \ln (s)]^{(1/p)}
 \label{froissart}
\end{equation}
apart from a  possible   very slow  $s$-dependence  from ${\bar \Lambda}^2$. 
 The same result  is also obtained using the saddle point method.

This indicates that the Froissart bound is saturated if $p=1/2$. We shall now 
show that,  in our model, analyticity demands $p>1/2$ and thus that, 
no matter how fast the mini-jet cross-section may grow with energy,  
the Froissart bound is always satisfied.

The requirement  that $p> 1/2$ follows from analyticity arguments in the 
complex $z_s$-plane, where $z_s=\cos\theta_s=1+2t/(s-4m^2)
\rightarrow 1+2t/s$
 is the cosine of the s-channel scattering angle for the equal mass case. 
Basically, this limitation comes from asking that  for large values of the 
impact parameter space $b$, the eikonal be such to decrease at least like an 
exponential, i.e. to have at least 
\begin{equation}
{\cal I} m\chi(b,s)\rightarrow e^{-b\sqrt{t_0}}
\end{equation}
where $t_0> 0$  is  the boundary of the Lehmann ellipse on the real axis in 
the $z_s$ plane.
To see  the argument, consider the elastic scattering amplitude for a process 
$a+b\rightarrow a+b$, normalized in such a way that
\begin{equation}
\sigma_{tot}=2\pi {\cal I}m F(s,t=0)
\end{equation}
with 
\begin{equation}
F(s,t)= i\int d^2{\bf b}   [1-e^{i\chi(b,s)}]J_0(b\sqrt{-t})
\end{equation}

While  in the s-channel physical region, and for equal mass particles, 
$s\ge 4m^2$ and $t\le 0$,    for the Lehmann ellipse  $F(s,t)$ is analytic 
for $t\le t_0\le \mu^2$ where $\mu$ is some hadronic mass (e.g. twice the 
pion mass, as the smallest mass exchanged in the $t$-channel). The actual 
value is unimportant, what is needed is  $t_0> 0$. Then, for  $0<t \le t_0$ 
the argument of $J_0$ becomes imaginary so that   
\begin{equation}
F(s,t>0)=i \int d^2{\bf b} [1-e^{i\chi(b,s)}] I_0(b\sqrt{t})
\end{equation}
where $I_0$ is the Bessel function of the second kind. The  asymptotic 
behaviour of the Bessel functions  is such that
$I_0(y)$ for $y$ real, grows exponentially as $y$ becomes large. 
The integral at the r.h.s. of the above equation has to exist up to values of 
$t=t_0$ and for fixed ($0\le t\le t_0)$. Since, for large $b$-values,
\begin{equation}
I_0(b\sqrt{t})\simeq e^{b\sqrt{t}},
\end{equation}
for the integrand to be finite  the imaginary part of the eikonal function, 
${\cal I} m\chi(b,s)$ must go to zero at least like an exponential, i.e. at 
least 
\begin{equation}
{\cal I} m\chi(b,s)\Rightarrow e^{-b\sqrt{t_0}}
 \end{equation}
 Now, let us return to the expression for ${\cal I} m\chi(b,s)$ from the BN 
model, where  ${\cal I} m\chi(b,s) \simeq e^{-h(b,s)}$
up to exponential accuracy. Using the large-$b$ behaviour of the function 
$h(b,s)$, derived in previous papers and reproduced in the previous section,  
we see that
\begin{equation}
{\cal I} m\chi(b,s)\simeq e^{-h(b,s)}\simeq e^{-(\frac{b}{b_0})^{2p}}
\end{equation}
whence follows that $p>1/2$.
\section{\label{sec:scales}About scales and parameters in the BN model}
 The model we have described contains different scales.
As mentioned earlier, in this paper we only study the rising part of the 
cross-section, whose behaviour  is proposed  to  come from  processes for 
which the outgoing partons have transverse momentum 
$p_t\ge p_{tmin}\simeq  {\cal O} (1 \ GeV) $. 
  In our model, $p_{tmin}$ is  the  scale which separates  perturbative 
scattering processes  from  everything else. 
  Soft gluon emission introduces two more scales,   namely $\Lambda$ and 
$q_{max}$. The latter is the maximum transverse momentum in the integral  for 
soft gluon emission, it is of order $p_{tmin}$ and plays the role of the 
energy scale $E$ which appeared in QED radiative correction factors.  
Thus soft gluons satisfy the condition 
\begin{equation}
k_t \le q_{max} 
 \simeq {\cal O} (p_{tmin}),
\end{equation}
 since most of the parton-parton cross-section is peaked at $p_t=p_{tmin}$.
  The next scale $\Lambda\simeq {\cal O}(\Lambda_{QCD})$ separates the 
region of ultra soft gluons from the rest.  This region was originally  
neglected, on the basis that gluons with $|k_\perp|\ll \Lambda$ would see 
the hadron as a point-like object \cite{lipatov} and such emissions would 
have a small probability, because of colour screening.
 This argument is appealing, and similar to the one mentioned in 
Sect.~\ref{sec:resum}, but in our opinion, there is no compelling 
theoretical reason to assume that ultrasoft gluon emission in high energy 
reactions has low probability. This argument     could be 
applied    to an isolated hadron, but not to 
  high energy hadronic scattering described through the scattering of 
partons, where
soft gluon emission is stimulated by QCD interactions.  It is through this 
interaction that we can expect 
the transition 
between hadrons and quarks to arise. 
A singularity in the infrared region 
would 
indeed provide a cut-off to separate quarks from hadrons and lead to such 
transition.
 This is the rationale behind going into the zero momentum region, 
and enter it with a singular confining coupling between ultra-soft gluons and 
the quark current. In our model, these ultra soft gluons are important for 
the extremely large  impact parameter values, which enter  eikonal 
formulations  of the total cross-section at very large energy. For processes 
where such large-$b$ values do not play a role, this region  may be  
irrelevant, though. 

The ultra soft gluon 
 distribution which we have introduced  depends on the  parameter $p$ 
which regulates the infrared region of the soft gluon integral of 
Eq.~(\ref{hbq}). 
 In 
\cite{GGPS}, we have used the value $p=3/4$ 
but this  is a phenomenological value 
 which was obtained after performing various averages over the PDFs.  
A determination of the actual value of $p$ is 
beyond the scope of this paper, except for the fact that
$p<1$ for the soft gluon integral to converge and $p>1/2$ for analyticity of 
the scattering amplitude.  

One can evaluate the coefficient of the $(\ln{s})^{1/p}$ term in Eq. (\ref{froissart}) as given  by
\begin{equation}
\label{con}
C={{2\pi}\over{m^2_\pi}}({{m_\pi}\over{\Lambda}})^2
({{ 27 (1-p)^2\varepsilon}\over{4[1+(1-p) \ln (2q_{max}/\Lambda)]}}
)^{1/p}
\end{equation}
for  $N_f=3$, as appropriate for the total cross-section limit.
In the approximation in which 
 the $q_{max}$  term is neglected, Eq.~(\ref{con})
gives $\sigma_T \approx \pi/m_\pi^2 \ln^2 {s}$  for $p=1/2$, 
$\Lambda=100\ MeV$ and $\varepsilon =0.3$.   
   However,   $q_{max}$, which  provides an extra dynamical scale,   cannot, 
in general, be neglected: using Eq.~(\ref{con}) with $q_{max}\simeq 1\ GeV$, 
the above equation gives  $C\simeq {\cal O}( 0.1) \pi/m_\pi^2$, namely a 
constant which is one order of magnitude smaller than  
 in the case of the actual Froissart bound \cite{martinold}.

 We note that in the eikonal model as used here, one can derive the limit 
for the inelastic cross-section \cite{amartin} following  the same steps as
  above
obtaining a constant reduced by a factor 2, as 
in a black disk model.
\section*{Conclusions}
Using an eikonal mini-jet model for the total cross-section we have shown
 how soft gluon $k_t$-resummation in the IR region 
can  reduce the strong power-like rise due to the  minijet cross-section.
  We have found that   this  model   will always satisfy the bound
 $\sigma_{total} \le \ln^2 s$ provided 
 that  the infrared behaviour  of $\alpha_s$  reflect
 a  rising one gluon exchange potential: for a potential rising like $r^{2p-1}$ the total cross-section is  limited by an asymptotic behaviour $\simeq 
(\ln\  s)^{1/p}$, with $1/2<p<1$. This establishes the connection, in our BN model, between confinement  and  the satisfaction of limitations imposed by the Froissart bound.
\section*{Acknowledgments}
G.P.
 acknowledges enlightening discussions with E. Lomon, R. Ferrari and
  M. Procura, 
 and thanks the MIT Center for Theoretical Physics
 for hospitality.
This work was partially supported by the Department of Science and 
Technology, India, under the J.C. Bose fellowship, and   
by MEC (FPA2006-\-05294) and  Junta
de Andaluc\'\i a (FQM 101 and FQM 437). 

\end{document}